\documentstyle[12pt,aaspp4]{article}

\def\vx{{\bf x}}
\def\vv{{\bf v}}
\def\vbeta{{\bf\beta}}
\def\phiext{\Phi_{ext}}
\def\cs{c_s}
\def\cm{{\cal M}}
\def\cf{{\cal F}}
\def\sh{{\cal S_H}}

\def\pd#1#2{{\partial{#1}}\over {\partial{#2}}}
\def\p2d#1#2{{\partial^2{#1}}\over {\partial{#2}^2}}
\def\bnabla{\hbox{{\tensy\char114}}}

\def\km{{\rm km}}

\lefthead{Ostriker }
\righthead{  }

\begin{document}

\title{Dynamical Friction in a Gaseous Medium}

\author{Eve C. Ostriker}
\affil{Department of Astronomy, University of Maryland \\
College Park, MD 20742-2421}

\begin{abstract}

Using time-dependent linear perturbation theory, we evaluate the
dynamical friction force on a massive perturber $M_p$ traveling at
velocity $\bf V$ through a uniform gaseous medium of density
$\rho_0$ and sound speed $\cs$.  This drag force acts in the
direction $-\hat V$, and arises from the gravitational attraction
between the perturber and its wake in the ambient medium.  For
supersonic motion ($\cm\equiv V/\cs>1$), the enhanced-density wake
is confined to the Mach cone trailing the perturber; for subsonic
motion ($\cm<1$), the wake is confined to a sphere of radius $\cs t$
centered a distance $V t$ behind the perturber.  Inside the wake,
surfaces of constant density are hyperboloids or oblate spheroids
for supersonic or subsonic perturbers, respectively, with the density 
maximal nearest the perturber.  The dynamical drag force has the form
$F_{df}= - I\times 4\pi (G M_p)^2\rho_0/V^2$.  We evaluate $I$ analytically;
its limits are $I\rightarrow \cm^3/3$ for $\cm <<1$ and 
$I\rightarrow \ln(Vt/r_{min})$ for $\cm>>1$.  We compare our results to the
Chandrasekhar formula for dynamical friction in a collisionless medium,
noting that the gaseous drag is generally more efficient when $\cm>1$ but
less efficient when $\cm<1$.  To allow simple estimates of orbit evolution
in a gaseous protogalaxy or proto-star cluster, we use our formulae to 
evaluate the decay times of a (supersonic) perturber on a near-circular
orbit in an isothermal $\rho\propto r^{-2}$ halo, and of a (subsonic) 
perturber on a near-circular orbit in a constant-density core.  We also
mention the relevance of our calculations to protoplanet migration in a 
circumstellar nebula.
\end{abstract}

\section{Introduction}

The process of dynamical friction, defined as momentum loss by a
massive moving object due to its gravitational interaction with its
own gravitationally-induced wake, arises in many astronomical systems.
Examples of systems in which such effects are well known to be
important include stars in clusters or galaxies, galaxies in galaxy
clusters, and binary star cores in the common envelope phase of
evolution.  In the first two examples, the surrounding background
medium in general consists of a combination of collisionless matter
(stars, galaxies, dark matter) and gas, while in the third example the
surrounding medium is entirely gaseous.  As a corollary to the
dynamical friction process, in all these cases the background medium
is heated at an equal and opposite rate to the energy lost by the
perturber.

The analytic theory for the gravitational drag in collisionless systems was 
developed by \cite{cha43}, and over the decades since 
has enjoyed widespread theoretical application, extensive verification by 
numerical experiments, and well-documented embodiment in observed astronomical
systems.  The variety of important consequences of gravitational drag in 
collisionless astronomical systems includes mass segregation in star clusters,
sinking satellites in dark matter galaxy halos, orbital decay of binary
supermassive black holes after galaxy mergers, etc.;  see e.g. \cite{bin87}.

Less well-developed is the corresponding theory of dynamical friction
in a gaseous (i.e., collisional) medium.\footnote{Instead, most
studies of the gravitational interaction between a moving massive
body and the surrounding gaseous medium have focused on the problem
of accretion, following on the analysis of \cite{hoy39} and
\cite{bon44}, and the early numerical work of \cite{hun71}.} For
supersonic motion, analytic linear-theory estimates of the
gravitational drag under assumption of a steady state were obtained by
\cite{dok64}, \cite{rud71}, and \cite{rep80}.  The resulting estimates
for the drag force in the steady supersonic case take the form
\begin{equation}\label{ssdrag}
F_{SS}= 
{4\pi (G M_p)^2 \rho_0 \over V^2} \ln\left({r_{max}\over r_{min}}\right)
\end{equation}
where $r_{max}$ and $r_{min}$ correspond respectively to the effective
linear sizes of the surrounding medium and the perturbing object,
similar to the drag formula obtained for a collisionless medium.
Although there is some ambiguity in the definition of $r_{max}$ and
$r_{min}$, the estimate (\ref{ssdrag}) appears consistent with
calculations of the gravitational drag that are obtained as a
by-product of numerical hydrodynamic investigations focused on the
Bondi-Hoyle-Lyttleton accretion problem -- see, e.g. \cite{shi85},
\cite{sha93}, and Ruffert and collaborators (see \cite{ruf96} and
references therein).

For steady-state, subsonic motion of the perturber, 
the front-back symmetry of the perturbed
density distribution about the perturber led \cite{rep80} to 
argue that gravitational drag is absent in the subsonic, inviscid case.  
Considering
that the drag force for the supersonic case increases proportional to 
$V^{-2}$ with decreasing perturber speed, $V$, it seems counterintuitive 
for the dynamical drag to become exactly zero when $V$ becomes infinitesimally
smaller than the sound speed.  In this paper, we reconsider the 
linear-theory drag as a time-dependent rather than steady-state problem, and
arrive instead at a nonzero value for the dynamical friction for the
subsonic case, while still verifying that the drag force is maximized for
perturbers with $V\approx c_s$.
In \S2, we derive results for the perturbed density distributions created by,
and dynamical drag on, a massive perturber on a constant-velocity trajectory
through a uniform, infinite medium.  In \S 3 we relate our results to the 
classical (collisionless) dynamical friction formula, and briefly consider
applications to a few astronomical systems.

\section{Analysis}

\subsection{Wave Equation and Formal Solution using Green Functions}

We begin with the linearized equations for the perturbed density
$\rho\equiv \rho_0[1+\alpha(\vx,t)]$ and velocity 
$\vv\equiv c_s \vbeta(\vx,t)$  
of an adiabatic gaseous medium that is 
subject to an external gravitational potential $\phiext(\vx,t)$:
\begin{equation}\label{continuityeq}
{1\over c_s}{\pd{\alpha}{t}} + \bnabla\cdot \vbeta =0
\end{equation}
and
\begin{equation}\label{momentumeq}
{1\over c_s}{\pd{\vbeta}{t}} + \bnabla\alpha = -{1\over c_s^2}\bnabla\phiext,
\end{equation}
where $\rho_0$ is the unperturbed density, 
$\cs\equiv(\partial P_0/\partial \rho_0)^{1/2}$ is the sound speed,  and 
the perturbation amplitudes $\alpha, |\vbeta|<<1$.  By substituting 
equation (\ref{continuityeq}) in the divergence of equation (\ref{momentumeq}),
we have 
\begin{equation}\label{waveeq}
\nabla^2\alpha -{1\over\cs^2} {\p2d{\alpha}{t}} = 
- {1\over \cs^2} \nabla^2 \phiext\equiv - 4\pi f(\vx,t)
\end{equation}
where $\rho_{ext}(\vx,t) \equiv \cs^2 f(\vx,t)$ is the mass density of the 
perturber.

In the absence of any disturbance prior to the action of $\phiext$,
the solution to equation (\ref{waveeq}) is found using the retarded
Green function $G^+$ for the 3D wave equation (\cite{jac75} \S 6.6) as
\begin{equation}
\alpha(\vx,t)=\int\int d^3 x' dt' 
{\delta[t'-(t-|\vx-\vx'|/\cs)] f(\vx',t')\over |\vx-\vx'|}
\end{equation}

\subsection{Perturbed Density Distributions for Constant-Velocity Perturbers}

We now specialize to the case of a point mass $M_p$ on a straight-line
trajectory with velocity $V\hat z$, passing at time $t=0$ through
$\vx=0$.  If ${\cal H}(t)$ describes the time over which the perturber
is active, then 
$f(\vx,t)=(G M_p/\cs^2) \delta(z-Vt) \delta(x)\delta(y){\cal H}(t)$.
We define $s\equiv z-Vt$ as the distance along the line of motion
relative to the perturber, $w\equiv z'-z$, and $\cm\equiv V/\cs$ as the
Mach number; $w>0 (<0)$ corresponds to the perturbation from a backward 
(forward) - propagating wave.  Then 
\begin{equation}\label{alphaintegral}
\alpha(\vx,t)={G M_p\over\cs^2} \int_{-\infty}^{\infty} dw
{\delta[w+s+\cm (R^2 + w^2)^{1/2}] {\cal H}((w+z)/V) \over 
(R^2 + w^2)^{1/2}}.
\end{equation}
Here $R=(x^2+y^2)^{1/2}$ is the cylindrical radius.

To evaluate the integral (\ref{alphaintegral}), one expands the 
argument of the $\delta-$ function about its possible roots
\begin{equation}\label{wpm}
w_\pm= {s\pm \cm [s^2 + R^2 (1-\cm^2)]^{1/2} \over \cm^2 -1};
\end{equation}
for $\cm <1$, only $w_+$ is a valid root, whereas for $\cm>1$ 
both roots are valid as long as $s<0$ and $|s|/R> (\cm^2-1)^{1/2}$,
and neither otherwise.  Using $\delta((w-w_\pm)A)=\delta(w-w_\pm)/|A|$
for $A=1+\cm w_\pm/(R^2+w_\pm^2)^{1/2}$ and substituting the solution
(\ref{wpm}), 
the result is
\begin{equation}\label{alphasum}
\alpha = {G M_p/\cs^2\over [s^2 + R^2 (1-\cm^2)]^{1/2}}
{\sum_{\rm roots\ w_0}} {\cal H}((z+w_0)/V).
\end{equation}

For a purely steady solution over all time, $\cal H$ is unity for all 
arguments, with the result that 
\begin{equation}\label{alphas}
\alpha_S= {G M_p/\cs^2\over [s^2 + R^2 (1-\cm^2)]^{1/2}} \times
\left\{ \begin{array}{ll}
1 & \mbox{if $\cm<1$} \\
2 & \mbox{if $\cm>1$ and $s/R<-(\cm^2-1)^{1/2}$} \\
0 & \mbox{otherwise}
\end{array}
\right. 
\end{equation}
This confirms the analytic, linear-theory results of previous authors
for completely steady flow created by a point mass on a straight-line,
constant-speed trajectory: (a) A subsonic perturber generates a
density distribution centered on the perturber at $s=0$, with contours
of constant density corresponding to similar ellipses in the $s-R$
plane with eccentricity $e=\cm$, and the short axis along the line of
motion of the perturber (i.e. the 3D density distribution consists of
concentric similar oblate spheroids).  The density contrast
$\rho_1/\rho_0\equiv\alpha$ is unity for an elliptical section with
semiminor axis $G M_p/\cs^2$; the linearization is sensible only
outside of this elliptical section.  This density distribution is a
generalization of the far-field limit of the hydrostatic envelope
$\rho/\rho_0=exp(G M_p/(\cs^2 r))$ that surrounds a stationary
perturber.\footnote{We note that for subsonic flow, the far-field
density enhancement from equation (\ref{alphas}) is within 10\% of
Hunt's (1971) $\cm=0.6$ numerical solution outside of $r=G
M_p/\cs^2$.  Also, the rms anisotropy of 7\% from the far-field
steady solution (\ref{alphas}) with $\cm=0.6$ is comparable to the
5\% mean anisotropy cited for Hunt's near-field numerical solution.}
(b) A supersonic perturber generates a density wake only within the
rear Mach cone of half-opening angle $\sin\theta=1/\cm$ defined by
$s/R<-(\cm^2-1)^{1/2}$; the surfaces of constant density within the
wake correspond to hyperbolae in the $s-R$ plane, with eccentricity
$e=\cm$.

Now consider the case where the perturber is ``turned on'' at $t=0$, so
that $\cal H$ is a Heaviside function.  For a subsonic perturber $\cm<1$ the
only root is $w_+$ (eq. \ref{wpm}).  Algebraic calculation shows that 
$z+w_+>0$ when $R^2+z^2<(c_st)^2$, for $\cm<1$.  Thus, the region of perturbation
is the sphere centered on the original position of the perturber,
within which a sound wave has traveled in time $t$.  Within this region of
perturbation, the density distribution has reached the value given by the
steady solution (eq. \ref{alphas});  outside the causal region for sound
waves, the density remains unperturbed.  

For a supersonic perturber, any
density disturbance must be confined within the rear Mach cone 
$s/R<-(\cm^2-1)^{1/2}$.  Algebraic calculation shows that within
the sphere $R^2+z^2<(c_st)^2$, $z+ w_+>0$ and $z+w_-<0$; hence
only $w_+$ contributes in equation (\ref{alphasum}).  Within the
Mach cone and to the right ($R<|z-\cm c_st|/(\cm^2-1)^{1/2}$, $z>c_st/\cm$) 
of this
sphere ($R^2+z^2>(c_st)^2$), both $z+w_+$ and $z+w_-$ are real and 
positive, hence contribute in eq. (\ref{alphasum}).  

The results for the perturbed density for this finite-time perturbation
are summarized as follows:
\footnote{These results appear to have been obtained previously by
\cite{jus90} via an alternative mathematical formalism;  the two calculations
serve as independent checks of the formulae.}
\begin{equation}\label{alpha}
\alpha(t)= {G M_p/\cs^2\over [s^2 + R^2 (1-\cm^2)]^{1/2}} \times
\left\{ \begin{array}{ll}
1 & \mbox{if $R^2+z^2<(c_st)^2$} \\
2 & \mbox{if $\cm>1$, $R^2+z^2>(c_st)^2$, $s/R<-(\cm^2-1)^{1/2}$, 
and $z>c_st/\cm$} \\
0 & \mbox{otherwise}
\end{array}
\right. 
\end{equation}
The region of perturbed density has the shape of a loaded ice-cream-cone 
dragged by its point by the perturber $M_p$;  only $w_+$ contributes in the 
ice-cream region (``region 1''), while both $w_\pm$ contribute in the cone 
region (``region 2'').  The cone shrinks in size as $\cm$ decreases for 
$\cm>1$, and is nonexistant for $\cm<1$.  Figures 1 and 2 show examples of 
the perturbed density distributions for subsonic and supersonic perturbers, 
respectively.  Because of the linear-theory assumptions made at the outset,
we note that equation (\ref{alpha}) is properly valid only for 
$\alpha <<1$, i.e. $s^2+R^2(1-\cm^2)>> (G M_p/\cs^2)^2$.

\begin{figure}
\figurenum{1}
\plotone{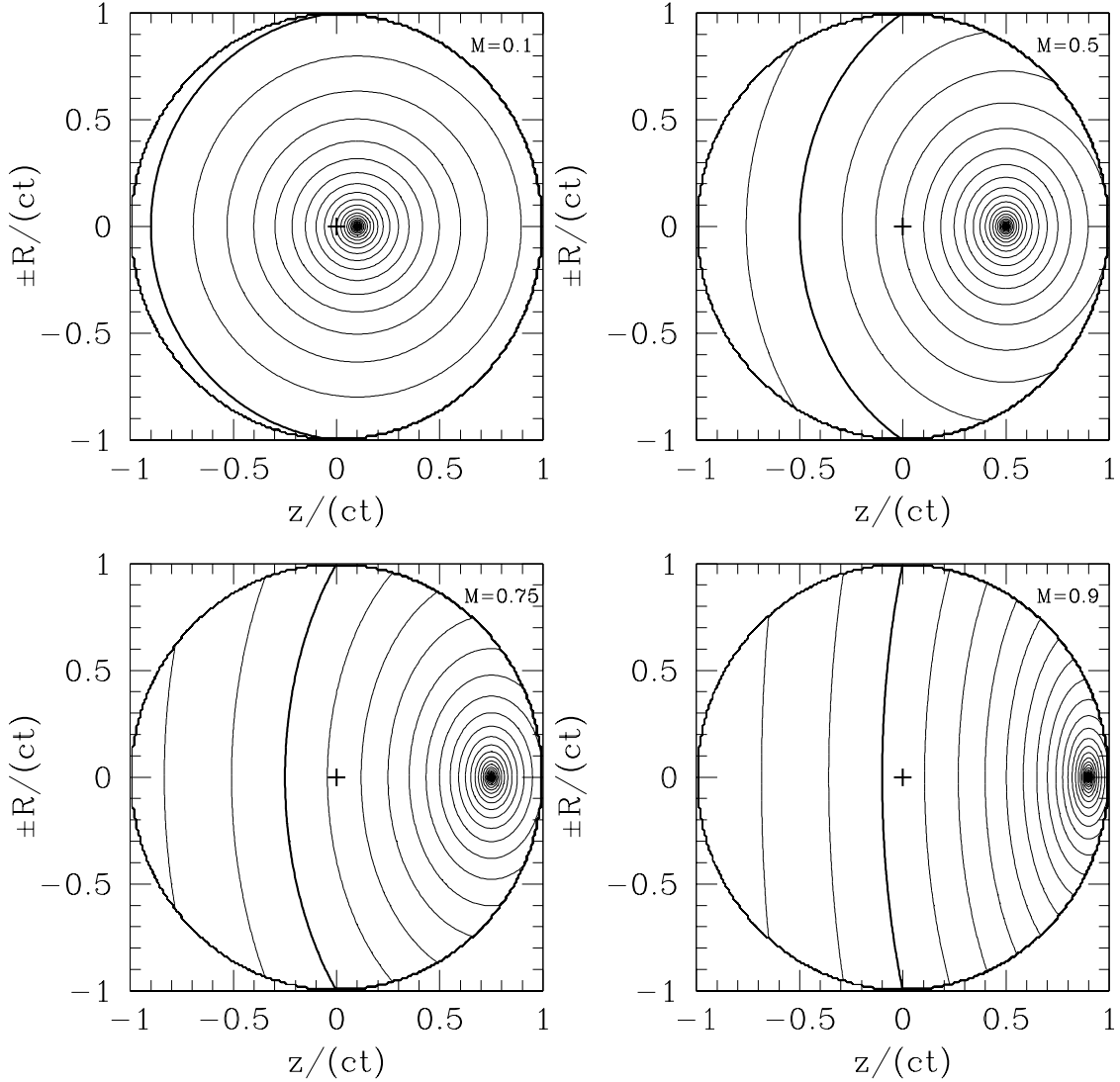}
\caption{Density perturbation profiles for subsonic perturbers with 
${\cal M}=0.1, 0.5, 0.75, 0.9$ (as indicated in upper right).
Contours show isosurfaces of 
$\log(\tilde\alpha)=\log(\alpha)-\log[G M_p/(t c_s^3)]$, 
in intervals of 0.1.  
Density increases toward the perturber; the heavy contour
indicates the surface with $\tilde\alpha=1$.  
The $\bf +$ symbol indicates the initial position of the 
perturber.}
\end{figure}

\begin{figure}
\figurenum{2}
\plotone{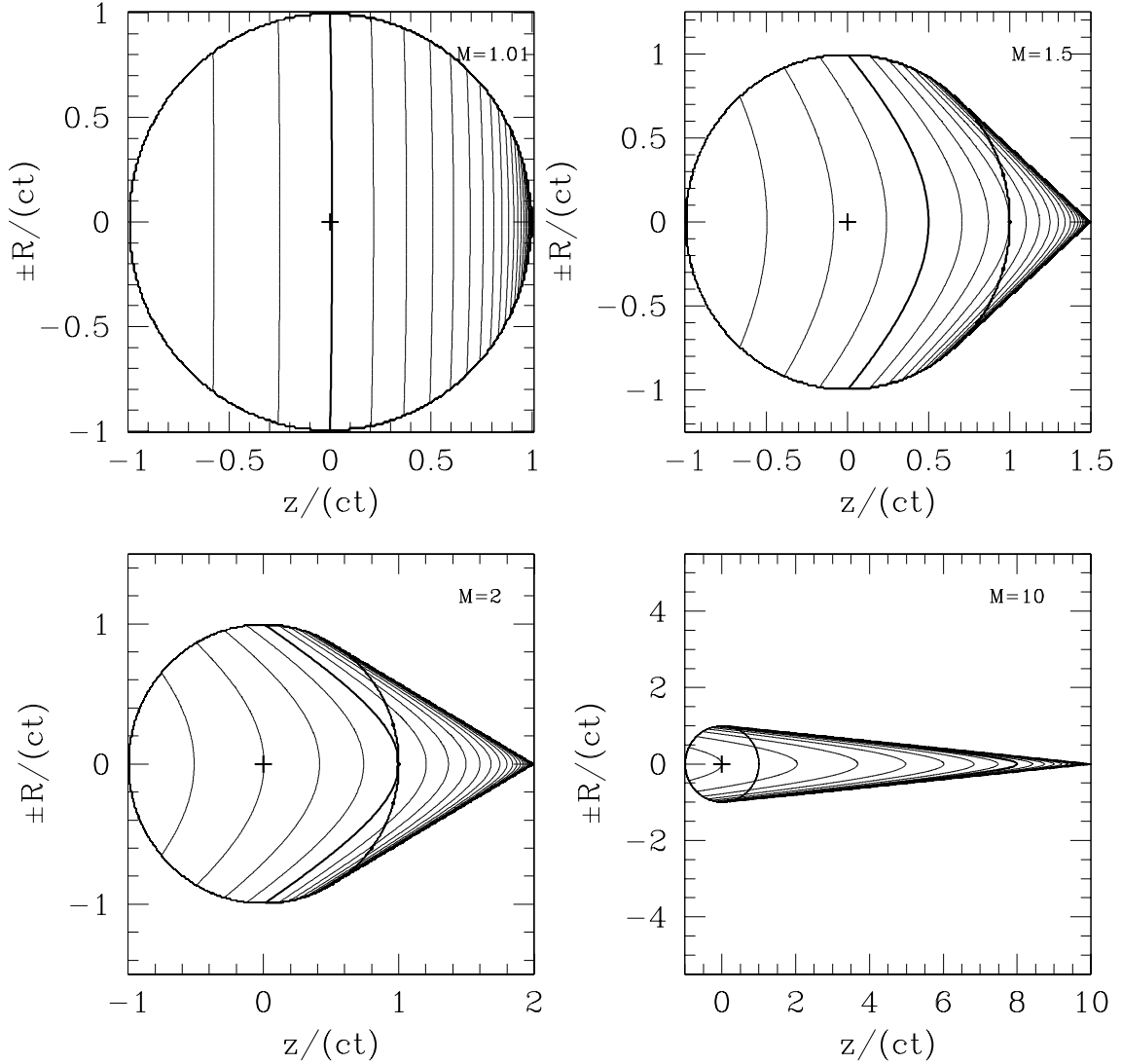}
\caption{Density perturbation profiles for supersonic perturbers with 
${\cal M}=1.01, 1.5, 2, 10$ (as indicated in upper right). 
Contours show isosurfaces of 
$\log(\tilde\alpha)=\log(\alpha)-\log[G M_p/(t c_s^3)]$, 
in intervals of 0.1.  
Density increases toward the perturber at the apex of the Mach cone.
The heavy contour
indicates the surface with $\tilde\alpha=1$.  There is a density jump 
with $\Delta \log(\alpha)=\log(2)=0.301$ at the surface $R^2+z^2=(c_st)^2$. 
The $\bf +$ symbol indicates the initial position of the 
perturber.}
\end{figure}

\subsection{Gravitational Drag Formulae} 

To compute the gravitational drag on the perturber, we need to evaluate
the gravitational force between the perturber and its wake,
\begin{equation}
F_{df}= 2\pi GM_p\rho_0\int\int ds dR R{\alpha(t) s\over (s^2 + R^2)^{3/2}}.
\end{equation} 
To perform the volume integral, it is convenient to transform to a spherical
polar coordinate system ($r,\theta$) centered on the massive perturber, with
$R\equiv r\sin\theta$ and $s\equiv r\cos\theta$.  Defining $\mu=\cos\theta$,
$x\equiv r/c_st$, we have 
\begin{equation}\label{dragdef}
F_{df} =-\cf I,\hskip1cm \cf \equiv {4 \pi (G M_p)^2 \rho_0 \over V^2},
\end{equation}
where 
\begin{equation}\label{dfint}
I=-{1 \over 2}\int {dx\over x} \int d\mu  
{\mu \cm^2 \sh \over (1-\cm^2 + \mu^2\cm^2)^{1/2}}.
\end{equation}
Here $\sh$ represents the sum in equation (\ref{alphasum}).

For the purely steady-state density perturbation given in equation
(\ref{alphas}), in the subsonic case $\sh=1$ everywhere in space and 
front-back antisymmetry in the angle integral (due to the symmetric 
spheroidal density distribution) argues that there is zero net force on the
perturber.  For the steady-state, supersonic case, $\sh =2$ for all 
$\mu$ between $-1$ and 
$\mu_M \equiv-\sqrt{\cm^2-1}/\cm$ (the boundary of the Mach cone),
so that $I=\int dx/x\equiv \ln(r_{max}/r_{min})$.  This is consistent with 
previous results, and yields an identical formula to
that representing the dynamical friction force in a collisionless medium when 
$V$ is much larger than the background particle velocity dispersion (e.g. 
\cite{bin87}, eq. 7-18). 

For the finite-time case, based on the perturbed density distribution in
equation (\ref{alpha}), there is nonzero contribution to the integral $I$ 
only from a finite region in space: region 1, in which $\sh=1$, and region
2, in which $\sh=2$.  Region 2 is nonexistant for subsonic perturbers.  
For the subsonic case, the perturber is surrounded by a concentric 
distribution of similar ellipsoids, and is displaced by $Vt$ forward 
from the center of the sonic sphere (radius $c_st$) surrounding its initial 
position.  The nearby, complete ellipsoids exert no net force on the 
perturber, but the larger, cut-off ones with semiminor axes between $(\cs-V)t$ 
and $(\cs+V)t$ lag behind the perturber (see figure 1) 
and exert a gravitational drag.  Thus, 
the radial integral in equation (\ref{dfint}) has upper/lower limits 
$x=1\pm \cm$, and the angular integral has limits 
$\mu=-1,\mu_C$ with $\mu_C\equiv (1-\cm^2-x^2)/(2 x\cm)$.  The integrals are straightforward; the 
result is
\begin{equation}\label{Isub}
I_{subsonic}= {1\over 2} \ln\left({1+\cm \over 1- \cm}\right) - \cm .
\end{equation}
The implicit assumptions in deriving this equation are that 
$(\cs-V)t$ exceeds the effective size of the perturber ($r_{min}$), and that
$(\cs+V)t$ is smaller than the effective size of the surrounding gaseous 
medium ($r_{max}$).  Under these conditions, the dynamical drag is 
time-independent, and non-zero.  The steady-state result that zero net force
results from the front-back symmetry of the density distribution is misleading;
because of the long-range nature of the Coulomb potential, the total drag 
force at any time depends on the unchanging ratio 
$(c_st+vt)/(c_st-vt)=(1 +\cm)/(1-\cm)$ 
between the semiminor axes of the furthest and closest perturbing 
{\it partial} 
spheroids.  The gravitational drag is always dominated by the far-field, 
and at any time the perturber is located ahead of center of the sonic sphere. 
Physically, we can associate the energy loss arising from the drag force
with the rate at which the expanding sound wave does work on the background 
medium.  In the limit of a very slow perturber $\cm<<1$, 
$I_{subsonic}\rightarrow \cm^3/3$, so that the drag force is proportional to
the perturber's speed $V$.  

For the supersonic case, the whole of the perturbed density distribution
lags the perturber.  The angular integration limits are $\mu=-1,\mu_\cm$
for $x=r_{min}/(c_st)$ to $\sqrt{\cm^2-1}$, and $\mu=-1,\mu_C$ for 
$x=\sqrt{\cm^2-1}$ to $ \cm+1$; $\sh$ takes on values 2 and 1 in regions 2, 1.
The result of the integration is
\begin{equation}\label{Isuper}
I_{supersonic}={1\over 2}\ln\left({\cm + 1\over \cm -1 }\right)+
\ln\left({\cm -1 \over r_{min}/c_st}\right)
={1\over 2}\ln\left(1-{1\over\cm^2}\right) + 
\ln\left({Vt\over r_{min}}\right).
\end{equation}
We have assumed that $Vt-\cs t>r_{min}$, and that the effective size
of the background medium exceeds $Vt+c_st$ .  In the limit $\cm>>1$,
we have $I_{supersonic}\rightarrow \ln(Vt/r_{min})$; with
$Vt\rightarrow r_{max}$, this recovers the steady-state
result.\footnote{As pointed out by the referee Scott Tremaine, the
  notion that the maximum impact parameter increases as $Vt$ was
  earlier introduced by \cite{ost68} in a time-dependent analysis of
  stellar relaxation.}

In Figure 3, we plot the dynamical friction force as a function of the Mach
number for several values of $c_st/r_{min}$.

\begin{figure}
\figurenum{3}
\plotone{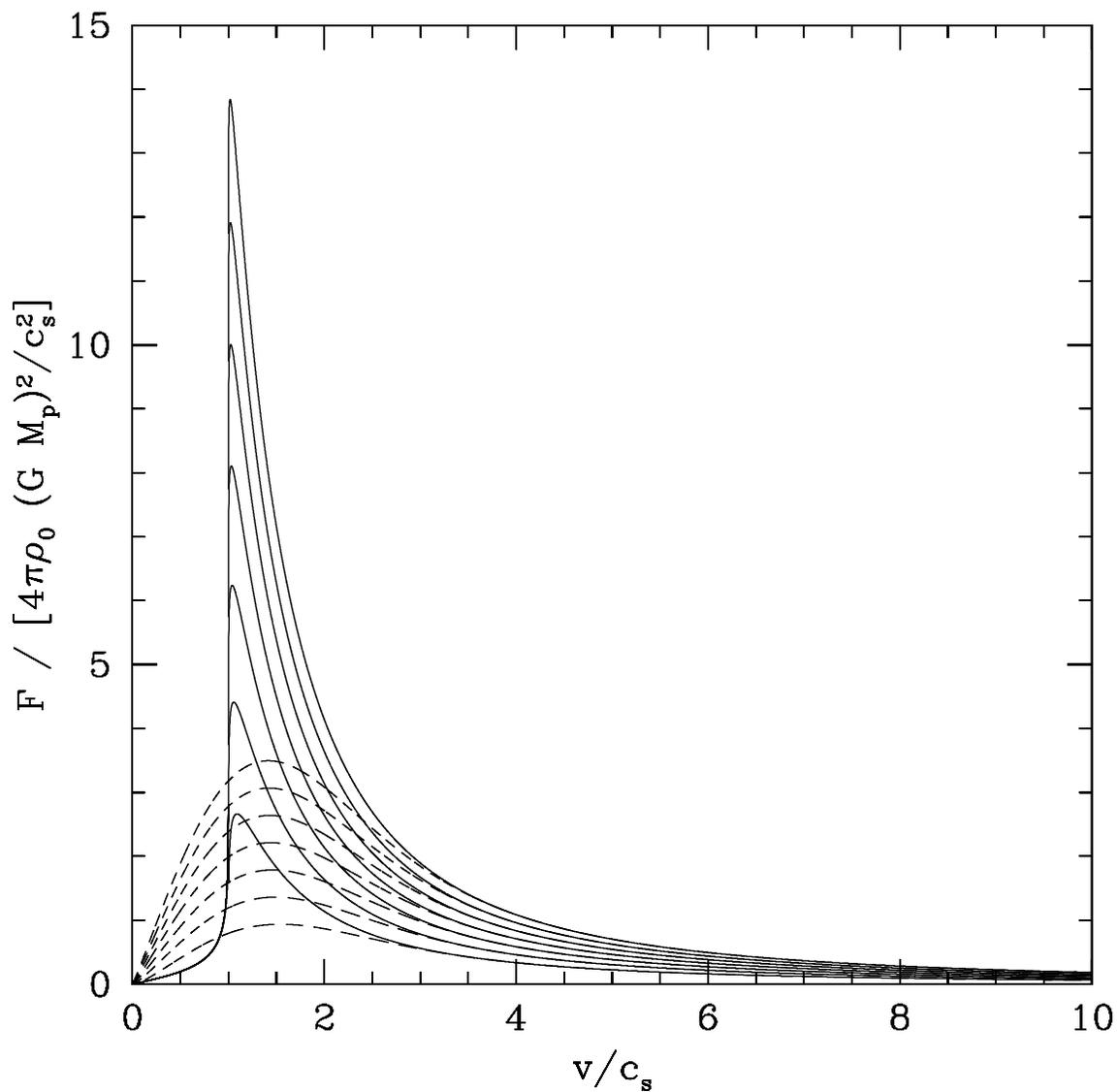}
\caption{{\it Solid curves:} Dynamical friction force in a gaseous medium 
as a function of Mach 
number ${\cal M}=V/c_s$.  Curves correspond to 
$\ln(c_st/r_{min})=4,6,8,...,16$.
{\it Dashed curves:} Corresponding dynamical friction force in a 
collisionless medium with particle velocity dispersion $\sigma=c_s$ and
$r_{max}\equiv Vt=\cm \cs t$.
}
\end{figure}

\section{Discussion}

The main formal result of this paper is the evaluation, in linear
perturbation theory, of the gravitational drag force $F_{df}$ on a
massive perturber $M_p$ moving on a straight-line trajectory through
an infinite, homogeneous, gaseous medium of density $\rho_0$ and sound
speed $\cs$.  Together with the definition in equation (\ref{dragdef}),
the expressions (\ref{Isub}) and (\ref{Isuper}) give the dynamical
friction (df) drag forces on subsonic and supersonic perturbers,
respectively.  Figure 3 presents the same results graphically, showing
how the drag force varies with the Mach number $\cm$ of the perturber,
and the time over which the perturber has been moving with fixed speed
$V=\cs \cm$.

For comparison, we have also included in Figure 3 the result for the
gravitational drag on a particle of mass $M_p$ moving through a 
collisionless medium with the same density $\rho_0$ as the gaseous medium
we have considered, and with a Maxwellian distribution of particle velocities
with $\sigma=c_s$.  From equation (7-18) of \cite{bin87}, the 
collisionless df drag force is 
given by expression (\ref{dragdef}) with 
\begin{equation}
I_{collisionless}=\ln\left({r_{max}\over r_{min}}\right) 
[{\rm erf}(X) - {2 X\over \sqrt{\pi}} e^{-X^2}],
\end{equation}
where $X\equiv V/(\sigma\sqrt{2})$.  From Figure 3, it is clear that
(a) for $\cm>>1$, the collisionless and gaseous df
forces are identical (as has been previously noted); (b) for $\cm<1$,
the drag force is generally larger in a collisionless medium than in a
gaseous medium, because in the latter case pressure forces create a
symmetric distribution in the background medium in the vicinity of the
perturber; (c) the functional form of the gaseous df drag is
much more sharply peaked near $\cm=1$ than it is for the 
collisionless df drag; perturbers moving at speeds near Mach 1 resonantly
interact with the pressure waves that they launch in the background
medium; (d) for a given value of $\ln(\Lambda)\equiv\ln(r_{max}/r_{min})$,
the peak value of the gaseous df force is much larger than the
corresponding peak value of the collisionless df force; at
$\cm\approx 1$, there is a factor of four difference in the force
between the two cases.  Below, we explore some potential consequences 
of these results in a variety of astronomical systems.

As a consequence of the stronger gaseous df force than collisionless
df drag force for supersonic motion, massive objects may make their
way more rapidly to the center of a star cluster or galaxy if they
arrive at the outer edge before, rather than after, the gas is turned
into stars.  For a particular example, we consider the decay of a
massive perturber's near-circular orbit in a spherical density
distribution with a singular isothermal sphere profile,
$\rho(r)=\cs^2/(2\pi G r^2)$ so $M(r)= 2\cs^2 r/G$ 
(here $\cs$ respectively denotes the sound
speed or velocity dispersion for a gaseous or stellar distribution).
For this density profile, the circular speed is constant, $V=\cs
\sqrt{2}$ .  By equating the rate of decrease of angular momentum 
$d(M_p V r)/dt$ to the torque $\tau_{df}= r F_{df}$, one finds that the 
time for the perturber's orbit to decay from $r_{init}$ to $r<<r_{init}$ is
given approximately by
\begin{equation}
{t_{df}(r_{init})\over t_{orb}(r_{init})}\approx
{M(r_{init})\over 4\pi M_p \ln\left({r\over r_{min}}\right)} 
\end{equation}
for gaseous df.  This time is a factor of $0.428$ shorter than the 
corresponding decay time under stellar df (cf. \cite{bin87}, eq. 7-25).
For galactic applications, the implication is that 
condensed objects that form early (e.g. globular clusters) could spiral into
the galactic center from a factor 1.5 further in a galaxy than would
be predicted by stellar df theory (\cite{tre75}), within the epoch over 
which the baryonic
distribution remains gaseous.  For star cluster applications, the shorter df 
time for gaseous distributions may help explain why young, embedded 
stellar clusters like the Orion Nebular Cluster (\cite{hil97}, \cite{hil98})
appear significantly more relaxed than expected from stellar df alone;  e.g.,
for the ONC, n-body simulations show that stellar df would require a
time a factor 3-4 longer than the best estimate of the cluster age to establish
the observed mass segregation (\cite{mar98}).  

As mentioned above (see Fig. 3), the gaseous df force is strongly depressed 
for subsonic perturbers.  For modeling the {\it global} evolution of combined 
star-gas systems in which the particle velocity dispersion and gas sound speed
are comparable, the strong cutoff of $F_{df}$ for subsonic perturbers 
implies that setting $F_{df}=0$ for $V<\cs$  (e.g. \cite{sai97}) should yield
realistic results.  However, in other circumstances it is interesting to 
enquire how the small, but nonzero, df drag on subsonic perturbers 
(computed in this paper) can affect their orbit evolution.

\begin{figure}
\figurenum{4}
\plotone{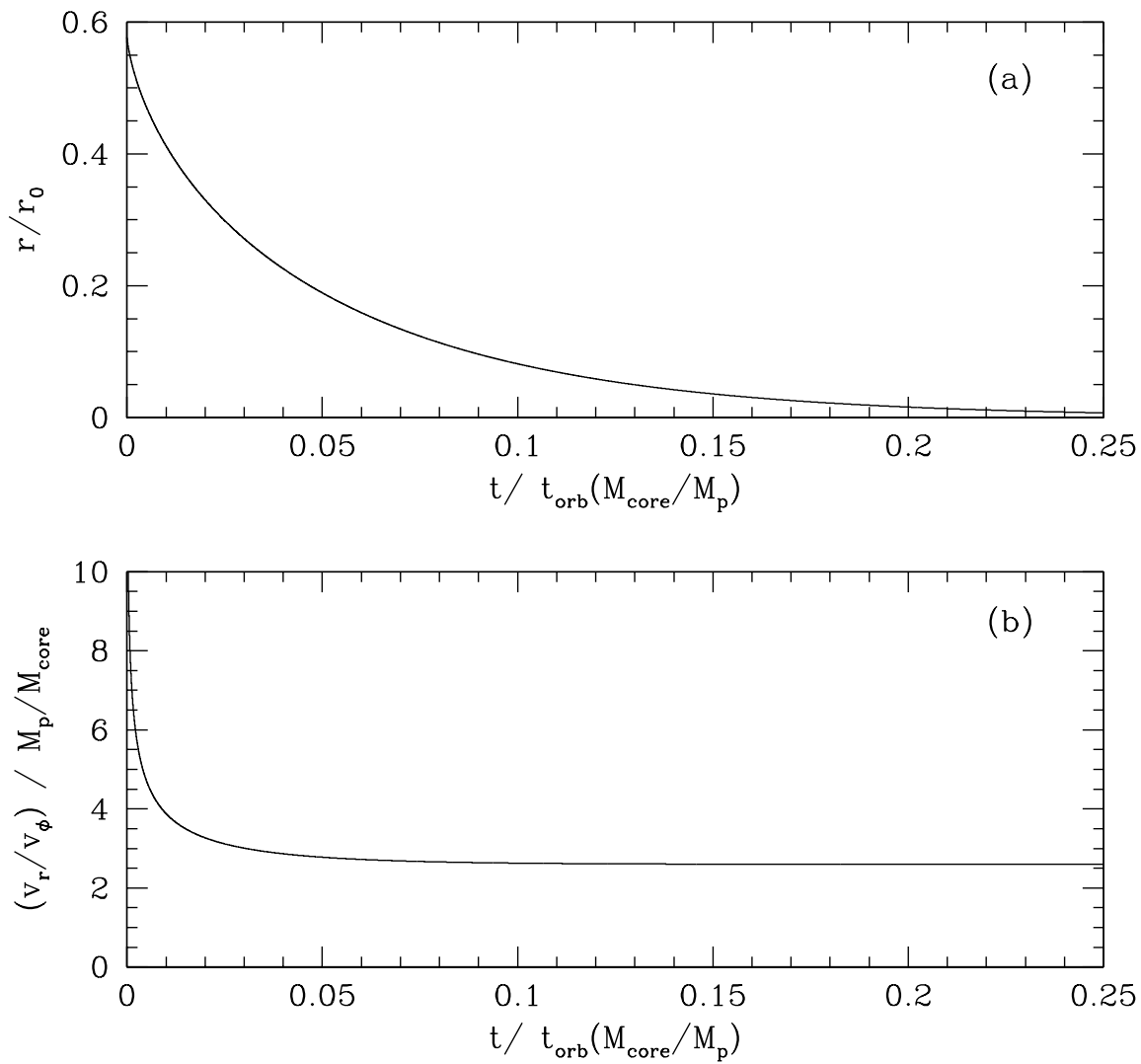}
\caption{(a) Decay in time of radial distance $r$ of massive perturber 
$M_p$ on near-circular orbit about the center of uniform-density core of 
mass $M_{core}$, radius $r_0$, and sound speed 
$c_s=(G M_{core}/(3 r_0))^{1/2}$;
(b) Radial-to-azimuthal velocity ratio for same situation as in (a).}
\end{figure}

\begin{figure}
\figurenum{5}
\plotone{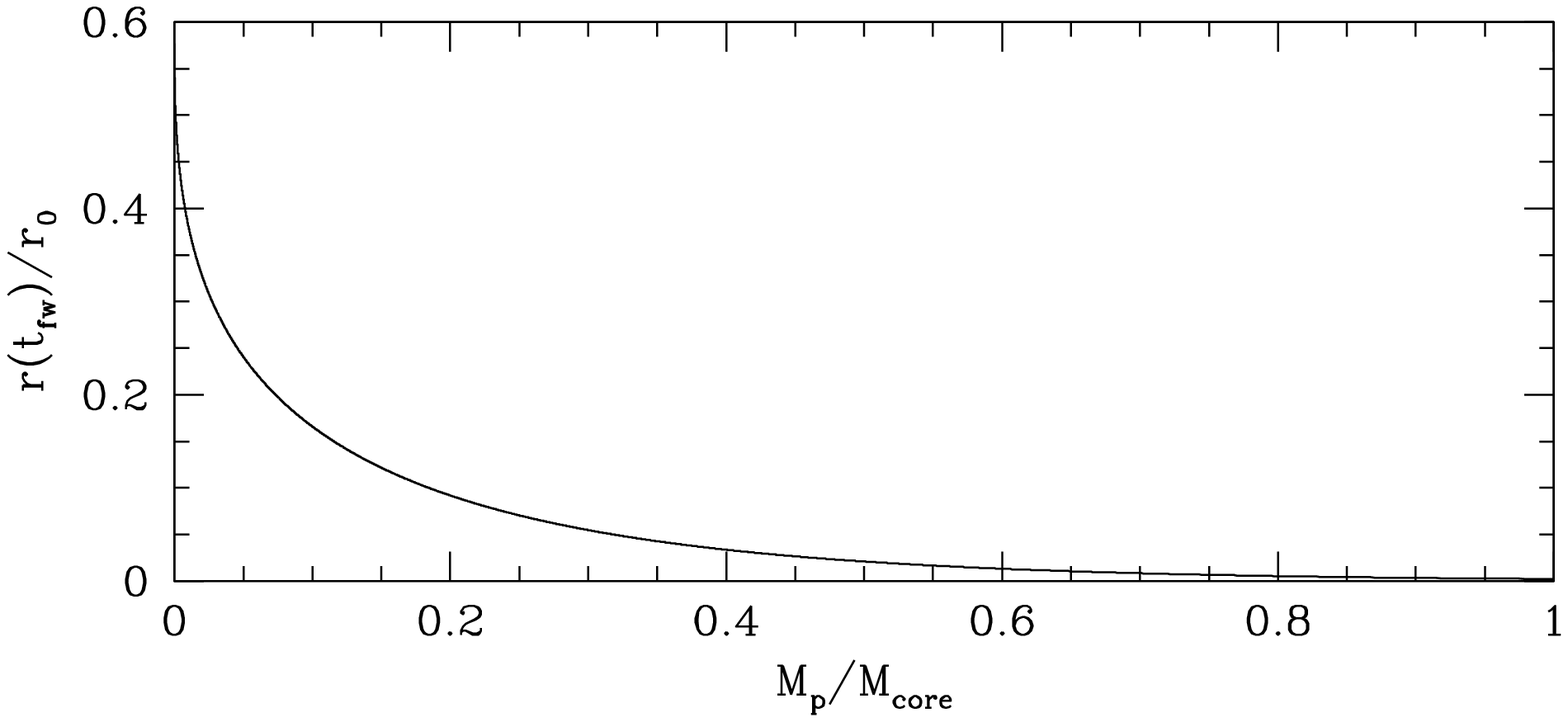}
\caption{For same situation as in Fig. 4, perturber's 
radial distance from the center at time when the forward wave disturbance
has propagated ahead of the perturber by $r_0$ (see text).}
\end{figure}

As a model problem, we consider the decay of a subsonically-moving mass $M_p$ 
on a near-circular orbit embedded within a constant-density gaseous sphere 
of radius $r_0$.  For constant background density, the angular orbit frequency
$\Omega=\sqrt{4\pi G\rho_0/3}$ and orbital period $t_{orb}=2\pi/\Omega$ are
independent of the distance $r$ from the center.  If this constant-density 
region represents the core of a nonsingular isothermal sphere with core radius
$r_0$ and sound speed $\cs$, then $r_0\Omega/\cs=\sqrt{3}$;  thus, the Mach 
number for a circular orbit at $r$ is $\cm=\sqrt{3}r/r_0$.  Just as above for
decay in a power-law density profile, we can compute the time for the 
perturber's orbit to decay from $r_{init}=r_0/\sqrt{3}$ to $r_f$ as 
\begin{equation}
{t(r_f)\over t_{orb}} = {M_{core}\over\pi 3^{5/2} M_p}
\int_{M_f={\sqrt{3}r_f\over r_0}}^1 {\cm d\cm\over I_{subsonic}(\cm)};
\end{equation}
the numerical coefficient equals $0.0204$.  
 
Figure 4a shows how the decay of the orbital radius depends on time
and on the mass of the perturber relative to the whole core.  In
Figure 4b, we verify that the assumption of a near-circular orbit is valid 
provided $M_p/M_{core}<<1$, since 
\begin{equation}
{v_r\over v_\varphi}= {3^{5/2} M_p\over 2 M_{core}} 
{I_{subsonic}(\cm)\over \cm^3},
\end{equation}
which has the limit $2.598 M_p/M_{core}$ for $\cm<<1$.  In arriving at the
results shown in Fig. 4, we have used equation (\ref{Isub}).  Its validity
depends, however, on the size of the uniform-density core exceeding that
of the perturbed-density region.  This assumption must fail, and the df drag
force consequently decrease, when the forward wave defining the 
disturbed-density region (cf. Fig. 1) reaches a distance $\sim r_0$ ahead
of the perturber.\footnote{Because the perturber follows a circular rather
than straight-line orbit, the df drag at late times should still be nonzero:
Since the direction of $\hat V$ changes by $\pi/2$ four times per orbit, the 
forward-wave propagation effectively ``restarts'' as well.  When $r<<r_0$, 
over each quarter orbit, each new expansion wave propagates to just $\sim r_0$;
over much of the quarter-orbit, an unbalanced trailing density enhancement 
will remain within the core.  Thus, we expect that the df drag will still be 
some fraction of the value found using eq. (\ref{Isub}), although a more
refined calculation is needed to predict the fraction quantitatively.}
The position $r(t_{fw})$ of the perturber at this time 
is plotted as a function of $M_p/M_{core}$ in Figure 5.  Based on this
figure, only perturbers of mass $M_p> 0.2 M_{core}$ are predicted to
reach within one-tenth of a core radius before the df drag decreases. 
For lower-mass perturbers, this implies that the decay of
orbits to very small radii may stall, if the df drag is sharply reduced 
after the time $t_{fw}$.  This result may 
have relevance for models of QSO evolution (\cite{sil98}) in which primordial
black holes are formed away from the centers of galaxies, later to be driven
there by df during mergers.  If, as assumed in this scenario, these events
occur before the advent of star formation, than the relevant df drag is the
gaseous df examined in this paper.  If the df drag becomes inefficient when
the orbit reaches the protogalaxy core and becomes subsonic, then
these massive black holes may have more time to grow by accretion before they
finally sink to the centers of their host galaxies.

Finally, we comment on the applicability of our results to the
interaction of protoplanets with a gaseous circumstellar nebula in
which they may grow.  Protoplanets orbit faster than the surrounding
gaseous nebula, with the Mach number of the drift speed $\cm\equiv
v_{rel}/\cs\approx \cs/(\Omega_K r) <<1$ (where $\Omega_K$ is the local
Kepler angular rotation frequency and $r$ is the protoplanet's
semimajor axis).  Drag forces occur as a result of this relative
motion, leading to inward radial migration of protoplanets as they
lose angular momentum to the surrounding nebula (e.g. \cite{war97} and 
references therein).  Using parameters
based on solid bodies in 
the protosolar nebula, it can be shown that the nominal df drag from equations
(\ref{dragdef}) and (\ref{Isub}), 
$F_{df}\approx (4\pi/3)(G M_p)^2\rho_0/(\cs\Omega_Kr)\approx 
(2\pi/3)(G M_p)^2 \Sigma/(\cs^2 r)$, exceeds the large-Reynolds number 
aerodynamic drag 
$F_{aer}\approx (1/4) v_{rel}^2 \pi R_p^2 \rho_0$ (\cite{lan87}, \S45)
for protoplanets of radius $R_p$ greater than a few $100\km$. 

It is sometime argued that gravitationally-enhanced drag on
protoplanets (cf. \cite{tak85}, \cite{oht88}, but note the different
scaling from our results) may enhance migration rates over those
predicted to arise from differential resonant torques (\cite{gol80})
between the protoplanet and the surrounding nebula.  In fact, provided
that the difference of gas orbits from Kepler orbits is included so
that the perturber does not corotate with the nebular gas at the same
radial distance, then the df drag is automatically incorporated in the
net resonant torque.  Indeed, calculations show that the net resonant
torque from a thin disk (e.g. \cite{kor93}, \cite{war97}), taking this
velocity difference into account, is of the same magnitude (and
scaling) as the df drag estimated above.  Thus, while the gaseous df
drag may have implications for the late stages of planet formation,
its effects are naturally incorporated within a full resonant
formalism.  Although such a calculation has not yet been performed in
three dimensions, it is reassuring that our simple estimate of the drag --
which neglects gradients of velocity, density, and temperature in the
disk -- and the estimates from 2D resonant torque differences -- which
neglect the disk thickness -- nevertheless yield comparable answers.

The author is grateful to the referee Scott Tremaine for helpful comments
on the manuscript.

\end{document}